# Component Based Software Development:
# A State of Art

N.Md Jubair Basha

Assistant Professor

IT Department, Muffakham Jah College of Engineering &
Technology, Hyderabad, INDIA

jubairbasha@mjcollege.ac.in

Dr. Salman Abdul Moiz

Professor

IT Department, MVSR Engineering College
Hyderabad, INDIA

Salman.abdul.moiz@ieee.org

*Abstract*— **one of the goals of Software design is to model a system in such a way that it is reused. Actively reusing designs or code allows taking advantage of the investment made on reusable components. However development of domain specific components and its impact on effort in terms of cost and time is still a challenging issue. The component based technology has transformed over a period of time from a simple component to the domain specific components. This paper presents a state of art of the drastic change in component technology from component engineering to domain engineering.**

*Keywords-Components; CBD,DE, product lines, Domain Specific Components frameworks.*

## I.    INTRODUCTION

Component based software development plays a vital role in increasing the productivity of an organization. There is a need for rich set of components in the repository which can be reused. In most of the projects, once the requirements are collected, the development activity starts from scratch. This may lead to overtime and over budget anomaly. If the existing component is reused rather than the developing the entire system from the scratch, not only the time is saved but quality product is produced.

An initial investment is required to start a software reuse process, but that investment pays for itself in a few reuses. In short, the development of a reuse process and repository produces a base of knowledge that improves the quality after each reuse cycle. This reduces the amount of development work required for future projects, and ultimately decreases the risk of new projects that are based on repository knowledge. It also helps in reducing the effort required for testing as the component in the repository is successfully tested.

The major advantages in using the domain specific components are

- Reduced Cost & Schedule: As the component is reused the cost and the time needed to develop the component is saved. If needed the component can be modified.

- Reduced Testing effort: As testing requires more than 60% of software development effort. Using domain specific component, testing effort is reduced.

- Enhanced Quality: As certification process is already completed for the developed component. As the component which is readily predicted to be of good quality.

Many organizations have developed their own domain specific components which act as an asset, so that they could be reused later. Even if the component is not reused as a mirror component, it can be modified. The effort required to modify a component is less compared to that developed from the scratch. However there is a need for an approach to identify and develop the domain specific components.

This paper presents state-of-the-art of the Component Based Software Development. Section –II presents Component Based Development (CBD) model. Several approaches [7] [16] of CBD are presented and compared. The Domain Engineering process [15] is presented with its working for the specific domain in Section – III. The purpose of DARE-COTS [1] tool is discussed along with the scope of product lines [11]. Section -IV presents the CBD-Arch-DE [14]. In Section –V, the Domain Specific Components Framework presents various component based frameworks, software architecture, components and performance of the systems [12, 10, 9, 8, 3] with the comparision of repositories and Section –VI includes the important issues and challenges for the CBSD. Section –VII concludes the paper.

## II.    COMPONENT BASED DEVELOPMENT MODEL

Spiral Models characteristics are involved in the Component Based Development (CBD) Model. The CBD Model consists of the applications from the grouped software components (called Classes) stated by [17].

The component model initiates with the identification of the candidate components. This is achieved by identifying the data to be changed by the application and the relevant algorithms that will be applied. For this the data and algorithms are encapsulated into the classes. The components created in the software projects are stored in repository. Once candidate components are identified, the repository is mined to check whether the desired components are present in the repository. If available, they are retrieved and are reused. If the component does not exist in the repository, it is engineered using the object-oriented methodology. The first iteration of the application to be build is composed of components retrieved from the repository and new components engineered to meet the novel needs of the particular application. Process Flow is reversed back to the spiral model and will ultimately continue the component assembly looping during subsequent passes through the component life cycle as shown in figure 1.

Software Reusability can be achieved by CBD model which is highly useful to the Software Engineers. Yourdon .E in [16] shows about the usage of software reusability by QSM Associates Inc., reports component assembly moves to a reduction in development life cycle, 84% reduction in project cost, and a productivity index of 26.2, compared to an industry norm of 16.9.With these results, the robustness of the component repository and CBD model provides many advantages to the software engineers

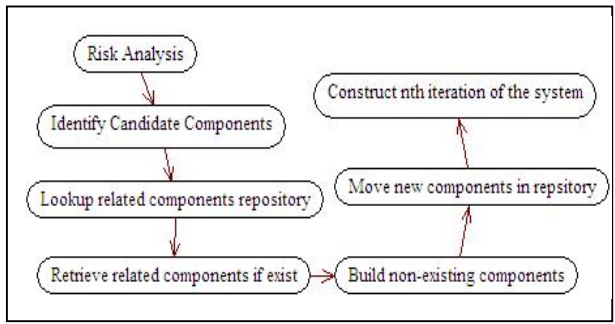

Figure 1: Component Based Development Model [17]

Sarbjeet Singh et al [7] surveyed the different concerns of reusability for component-based approach, metrics and models of software reuse. The research issue presented in this paper is the study and empirical validation of the proposed metrics for Component based system.

Component interface metrics have the potential to increase our understanding of the reusability of components. They are important because other sources of information relevant to reusability will often be unavailable as thirdparty components tend to be black-box, and because their automation can help provide a less biased and more accurate and efficient reusability analysis of components. The set of interface metrics developed here has demonstrated that measurement of component interfaces can provide useful and relevant information for the reusability analysis of components. The metrics can yield a significant body of information from interfaces, in a way that is more efficient than non automatable techniques. These metrics may give a better understanding of the properties of components' interfaces. The theory behind the metrics brought forth a reusability analysis of the test components that was generally consistent with expert knowledge of the test components. It follows that the metrics may be able to be used for reusability analysis of components with which the metrics practitioner is unfamiliar [22].

## III.    DOMAIN ENGINEERING

Domain Engineering (DE) is a process in which the reusable component is developed and organized and in which the architecture meeting the requirements of this domain is designed[24]. The "domain" refers to the functional areas covered by a group of application systems that have the same or similar software requirements [18].

Domain engineering process [15] is depicted in figure 2. DE consists of three main stages i.e. domain analysis, domain design and domain implementation. For Domain Analysis support, DARE-COTS tool is presented [1]. Initially, in a particular domain it is mandatory to get the universal and variable characteristics of group systems. By abstracting the characteristics, domain analysis model can be generated. Based on this model the domain specific software architecture can be designed and then reusable components will be generated and organized.

Thus, when developing a new system in new domain, we have to identify the system's requirements and specification as per the domain model, and can generate the new design as per the Domain Specific Software Architecture (DSSA), then select the particular components to assemble the new system. The process of developing an only single application system is called Application Engineering.

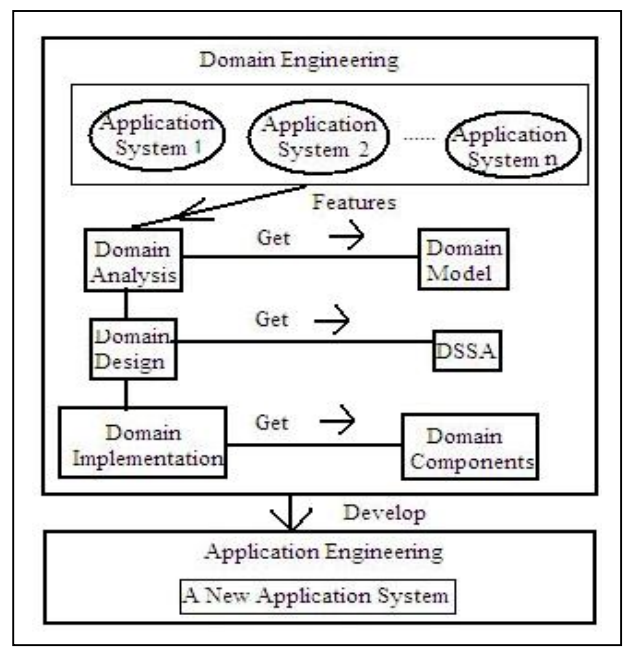

Figure 2. Process of Domain Engineering [15]

In [15] the method of domain engineering is described and the DSSA of Product Quality Tracking System has been presented. The discussion shows on how to develop an open and reusable product quality tracking system on the basis of domain engineering. The research shown in this article specifies the need of reusing the major functionality of the system when the application which is developed in a similar domain for which the components are available. Further research has to focus on a specific product quality tracking system using asserts and to perfect the component repository. Massiom et al [4] evaluated an application of domain analysis in a specific domain, i.e. production management and measured the results with some improvements to be done by integrating domain analysis method in standard development.

### A. Product Lines

The acronym for the domain engineering is product lines. The entire process of reusing domain knowledge is the production of new software systems. There are many definitions for software product lines. One of the widely accepted one's given by SEI is:

'..a set software intensive systems that share a common, managed set of features satisfying the specific needs of a particular market segment or mission and that are developed from a common set of core assets in a prescribed way'[6]. [11] Gives the clear introduction to the product lines, the related concepts related to product lines, technologies that apply to this method, success stories of organizations using this method. Product Line Engineering will be considered as a standard technique in Software Engineering. The research issues as par with the product lines is prioritize the assets if a group of products could be built through product line base methods, a tool support for all aspects of product line based software development is needed. With this the Product building will take less time, cost and posses more quality with respect to the software product lines.

### IV. CBD-ARCH-DE

Braga et al. [19] proposed a process-Odyssey-DE-that unifies the aspects of reuse and domain understanding as provided by DE methods and the representation of domain (in) variants as components provided by CBD methods. The set of activities referred to the domain design phase was not exhaustively addressed. CBD- Arch- DE (Component Based Development – Architecture –Domain Engineering) is a process proposed with the goal of detailing the states referring to architectural issues in design phase, which is the extension of Odyssey-DE process.

The domain design phase consists of two states: components generation and architecture generation which is proposed by Ana Paula et al.[14].In this paper a Grouping Criteria Tool has been proposed which is supported in the Odessy. This environment gives the available information of each criterion, suggestions of components groupings and related criteria based on the requirements. Further, future work can be done to improve the organization of the domain components architecture, its reuse in applications [14].

### V. DOMAIN SPECIFIC COMPONENT FRAMEWORKS

With the exponential increase in development of software systems in various domains, there is a dire need for development of Domain Specific Components frameworks (DSCF). Most of the software development activities followed today is agile in nature. It is believed that the DSCF's has been contributed from the patterns. Frederic et al [12] proposed Domain Components and identified patterns and developed a generic framework that itself provides the implementation of Domain Components semantics by evaluating domain-specific services with a unified approach. The proposed approach [12] conducted various case studies that lead to the different domains. The architectural patterns have been proposed together with the employed generative programming methods which includes the complexities of implementation of domain specific concerns. The open research issue in this paper remains consistent and symmetric approach to construction of containers needs to specify in the form of policies that will manage an important combinations of domain-specific services. Heiko et al. [13] focused the future directions for a specific end system domain (i.e. embedded or distributed systems) or for a specific technical domain area (e.g. EJB Systems), other domain-specific methods and models could improve the performance evaluation of component-based systems. For the particular business domains (e.g. accounting systems, retails systems, etc), patterns could be proposed so that special domain-specific languages and prediction methods can be designed. This will strongly support for the creation of prediction models and such that to cost is reduced for the model-driven performance analysis. One research issue is that it is difficult to the non-trivial software component applicable to the whole input domain. So, the suitable abstractions on for the input domain based on the equivalence classes needs to be proposed[20,21].The disjoint processes of software components productivity and grouping will posses a huge significant problem for all the stakeholders in the software components marketplaces. Much research needs to be carried out in this domain in order to provide more secured and efficient collection of software components [10].

Veikko et al [9] presented some of the issues to carry out research on the use of open source software in the context of both Original Software Component Manufacturing (OCM) and Commercial off-the-shelf (COTS) components. OCM process specifies the creation and improvement of software product-oriented view i.e. including both generic reference models and domain specific or product-specific platforms, middleware type integration techniques. Reference architecture can be used as a backbone for the exchange of components, products and platforms.

Sridaran el at[8] has analyzed different pattern based web applications for data handling and structuring issues. The future work direction specified in this paper is presenting an architecture that will be imported into the MVC framework without combining into the partition decision issues which is merely useful for the web applications.

### A. ISARE

Rizwan Ahmad el. At [3] proposed a framework on the aspects of software Architecture on quality attributes. It is mainly directed on reuse of available, proven techniques in efforts to maximize the results of a case study that **I**ntelligent **S**oftware **A**rchitecture **R**euse **E**nvironment (ISARE). It ensures the required level of quality requirements in the software architecture and can be automated.

The related work can be extended by reusing the existing proven Software Architecture evaluation techniques. Further, much work can be done by investigating and reusing the strengths of those important areas which have great impact on overall software quality concerns.

Another research direction is to strengthen the Software Architecture repository and automate the software architecture selection and evaluation methods for efficient and reliable systems. The main input in this repository is the set of architectural styles/artifacts and their comprehensive analysis against the set of quality attribute. Better utilizing the approaches for Software Architecture analysis could be the best possible option to move increase the reusability on existing data.

*B. TRUSTIE*

**T**rusted National Software **R**eso**U**rce **S**haring and Co-opera**T**eng EnvIronmEnt (TRUSTIE) [2] is a National 863 grand project. The purpose of this project is to construct a large-scale software production environment for trust worthy resource sharing and development cooperation. TRUSTIE is used as a command-line tool for component development. A new tool has to be developed to provide advantages of meta-model to model-editor generators. A research issue presents the close connection with Service Oriented Architecture [2].

*C. SE4SC*

Hao Chen et al [5] presented a search engine for the software components: SE4SC.The functionality of the proposed search engine is it collects the software component resources from the network and helps the software re-users to search the software components. This search engine not only support for the software components, but it can moves towards the re-users search request to the extracting engine for the software component library for accurate extraction.

The performance of SE4SC can be enhanced. Software Component Description Model (SCDM) can be improved by defining a more potential systematic classification schema and faceted classification schema. This search engine can be integrated with the extraction engines built in the component library and provide the better support for the re-users to search for the components which are having the equivalent requirements.

N Md Jubair Basha et al [25] has proposed a framework studio for effective component reusability which provides the selection of components from framework studioand generates source code based on the stakeholders needs. This framework studio is implemented in using swings which are integrated onto the NetBeans IDE which helps in faster generation of the source code.

Commercial reusable component repositories usually are integrated into a CASE environment . Currently, some major repositories (ASSET, PAL, and DSRS) begin to use web-based techniques to provide services. They are utilizing flat files written in HyperText Markup Language (HTML). Electronic Library Services and Applications (ELSA) has gone a step further by using the Multimedia Oriented Repository Environment (MORE). Web-based reuse is the trend of software component repositories supported by the government. To be a part of an integrated CASE environment is the trend of commercial software component repositories. Usually, the aim of the first one is to provide a service within a domain, organization, or area, such as ASSET for DoD, DSRS for DISA etc. This kind of repository is used in a wide scope. The aim of the second is to provide an integrated CASE environment for a software development organization. So, this kind of repository is generally a part of CASE environment and is used in a relatively narrow scope[23].

| Features | Web-Based | Integrated into CASE Environment | Security Control | Retrieval Methods |
|---|---|---|---|---|
| +1 Reuse Repository | | Y | Y | Browsing |
| SALMS | Y | Y | Y | Keywords |
| ASRR | | | Y | Keywords |
| The Universal Repository | | Y | Y | Browsing and Keywords |
| AIRS | | | Y | Facets Approach |
| RLT | | | Y | Keywords |
| HSTX Reuse Repository | Y | | Y | Keywords |
| DSRS | Y | | Y | Keywords |
| LID | | | Y | Keywords |
| I-CASE | | Y | Y | Keywords |
| MORE | Y | | Y | Keywords |
| ASSET | Y | Y | Y | Keywords |
| PAL | Y | | Y | Keywords |
| CAPS | | Y | | Browsing, Keywords, Profile & Signature Matching |
| Ada Library and Reuse Library (DISA) | Y | | Y | Browsing and Kewwords |

Figure 3: Comparision of Repositories

## VI.  ISSUES & CHALLENGES

Most of the software development applications are developed using an agile methodology.  This approach may or may not use the existing components. Even if several components are reused, an agile approach doesn't update the repository and the components become obsolete after certain time period. A repository of components is to be maintained for effective reuse of the components. The organizations usually function in multiple verticals. As such there is a need for domain engineering process which groups the related components. However this is possible only if a robust Domain Specific Components framework is available.

Several organizations maintain the domain specific components as their assets. However there are certain unsolved issues. The organizations have to modify their development process consisting of two sub processes i.e. software development for re-use and software development with re-use. As software components are not plug and play type, integrating them becomes a challenging issue but a better framework may resolve the integration issues.

Standardization of component development process is a challenging issue. There is a need for unified component development and unified component testing process. A component maturity model as similar to capability maturity model is needed. The key process areas of a component maturity model need to be identified and standardized to know the maturity of an organization following a component development approach.

Though there is a rich set of software metrics available but there is a dire need for research in developing the component metrics as the program volume measure, potential volume measure or software cyclomatic complexity measure has to be revived with respect to the component based software development.

As the cost estimation is an error prone task, several models to estimate the cost of a new component and the cost of a modified component can be thought of.

## VII.  CONCLUSION

Putting all these together, we opine that CBD-Arch-DE is the most promising branch of  CBSD. This is not only because their focal models are good at revealing system essential and have some desirable properties such as design as orientation, extensibility and reusability.  The comparision of repositories gives an idea for the best suitable to select from the available for the component and artifact extraction. Achieving a positive future we will require, however specific advances in Domain Specific Modelling, Domain Specific Components and Domain Specific Architectures. It is necessary to work a lot on Domain Specific Services as some any how only three metrics are discussed. By considering the above discussed issues and challenges it is possible to develop the domain specific components  and its impact on effort in terms of cost and time can be reduced by achieving reusability in the software development. Framework for the Component Reusability is needed so  that the different components can be extracted from the different domains from a particular repository. It is intended to develop the generic domain specific components by  proposing the framework for the above said and the work is extended   towards the development of a tool for the extraction of different components for multiple domains. In order to obtain this, we have followed the CBD-Arch-DE process which is considered as a better approach among the both CBSD and Domain Engineering. One of the approaches discussed in this paper is Software Architecture which is mainly considered for reuse. As our work mainly concerned with the reusability issue, so we have considered the Intelligent Software Architecture Reuse Environment. Given that in Domain Specificity could be a high level architecture, code generation, especially, non-structural code generation is still a research challenge. The component metrics can be considered and performance metrics for domain specific services can be identified as challenging research. further analysis is needed to determine the relative usefulness of the metrics, and to give metrics practitioners confidence in applying them. In particular, axiomatic analysis and further empirical analysis are desirable.

Modeling", Tsinghua University Press, Beijing, 2008.

**Authors**

**N Md Jubair Basha** received his B.Tech. (IT) and M.Tech (IT) from JNTUH,    Hyderabad He is presently working as Assistant Professor in Department of Information Technology, Muffakham Jah College of Engineering and Technology, Hyderabad, India. His research interest includes Component Based Software Development, Software Reusability, Reverse Engineering and Cryptography. He is an active member of IEEE and CSI.

**Dr. Salman Abdul Moiz** is working as a Professor in CSE/IT department at MVSR Engineering College, Hyderabad. He received B.Sc (Electronics) from Osmania University, MCA from Osmania University, M.Tech (CSE) from Osmania University, and M.Phil (CS) from Madurai Kamaraj University and Ph.D (CSE) from Osmania University.  He worked as Research Scientist at Centre for Development of Advanced Computing, Bangalore. He has published 33 papers in various National/International Conferences and Journals. His areas of interests include Mobile databases, Software Process Improvements; Component based software development & Disaster Recovery. He is an active member of IEEE, IETE and CSI.